\documentclass[conference]{IEEEtran}
\IEEEoverridecommandlockouts
\usepackage{cite}
\usepackage{amsmath,amssymb,amsfonts}
\usepackage{algorithmic}
\usepackage{graphicx}
\usepackage{textcomp}
\usepackage{xcolor}
\usepackage{url}
\usepackage{booktabs} 
\usepackage{hyperref} 
\def\BibTeX{{\rm B\kern-.05em{\sc i\kern-.025em b}\kern-.08em
    T\kern-.1667em\lower.7ex\hbox{E}\kern-.125emX}}
\begin{document}

\title{LHQ-SVC: Lightweight and High Quality Singing Voice Conversion Modeling\\
}






\author{
	\IEEEauthorblockN{
		Yubo Huang\IEEEauthorrefmark{1}\IEEEauthorrefmark{6},
        Xin Lai\IEEEauthorrefmark{1},
		Muyang Ye\IEEEauthorrefmark{2}, 
		Anran Zhu\IEEEauthorrefmark{3}, 
		Zixi Wang\IEEEauthorrefmark{3},
		Jingzehua Xu\IEEEauthorrefmark{4},
  		Shuai Zhang\IEEEauthorrefmark{5},\\
        Zhiyuan Zhou\IEEEauthorrefmark{3},
		and Weijie Niu\IEEEauthorrefmark{7}} 
	\IEEEauthorblockA{\IEEEauthorrefmark{1}School of Civil Engineering, Southwest Jiaotong University, China} 
	\IEEEauthorblockA{\IEEEauthorrefmark{2}SWJTU-Leeds Joint School, Southwest Jiaotong University, China}
	\IEEEauthorblockA{\IEEEauthorrefmark{3}School of Computing and Artificial Intelligence, Southwest Jiaotong University, China} 
	\IEEEauthorblockA{\IEEEauthorrefmark{4}Shenzhen International Graduate School, Tsinghua University, China}
    \IEEEauthorblockA{\IEEEauthorrefmark{5}Department of Data Science,
    New Jersey Institute of Technology, USA}
    \IEEEauthorblockA{\IEEEauthorrefmark{7}School of Economics and Management, Southwest Jiaotong University, China}
   \IEEEauthorblockA{\IEEEauthorrefmark{6}Zhenguan AI Lab, ZhenGuan Innovation (Shenzhen) Technology Co. Ltd, China}
}

\maketitle

\begin{abstract}
Singing Voice Conversion (SVC) has emerged as a significant subfield of Voice Conversion (VC), enabling the transformation of one singer's voice into another while preserving musical elements such as melody, rhythm, and timbre. Traditional SVC methods have limitations in terms of audio quality, data requirements, and computational complexity. In this paper, we propose LHQ-SVC, a lightweight, CPU-compatible model based on the SVC framework and diffusion model, designed to reduce model size and computational demand without sacrificing performance. We incorporate features to improve inference quality, and optimize for CPU execution by using performance tuning tools and parallel computing frameworks. Our experiments demonstrate that LHQ-SVC maintains competitive performance, with significant improvements in processing speed and efficiency across different devices. The results suggest that LHQ-SVC can meet real-time performance requirements, even in resource-constrained environments.
\end{abstract}


\section{Introduction}

Singing Voice Conversion (SVC) has emerged as a promising application in the music domain, offering new possibilities for music production, cover songs, and personalized voice customization. SVC involves not only the conversion of speech characteristics but also musical elements such as melody and rhythm, making it more challenging than traditional VC tasks. 

Traditional SVC methods have largely relied on generative models. Nose et al. \cite{nose2015} proposed an HMM-based singing synthesis technique that allows users to intuitively and continuously control the singing style and intensity of synthesized singing voices. Kobayashi et al. \cite{kobayashi2014} introduced a statistical conversion process based on GMM to estimate spectral features directly, addressing the degradation in voice quality during conversion. Sisman et al. \cite{sisman2019} were the first to attempt using GANs for SVC and proposed a framework named SINGAN, which achieved high-quality singing voice conversion without relying on automatic speech recognition systems.

Despite their advancements, these models face several limitations, including high data requirements, loss of audio quality, and the complexity of model training. In comparison to traditional statistical methods, deep learning has demonstrated superior performance by learning complex nonlinear mappings from large datasets, significantly improving the quality of converted voices and speaker similarity \cite{sisman2020}. Sun et al. \cite{sun2015} proposed a sequence conversion method based on DBLSTM-RNNs, utilizing LSTM to improve the naturalness and continuity of speech output in VC tasks through parallel training data. Xie et al. \cite{xie2016} introduced a DNN-based non-parallel training approach to VC, achieving effective conversion without parallel data.

Recently, diffusion models have emerged as a powerful generative technique in image and audio synthesis. These models learn data distribution by gradually introducing and removing noise, capturing subtle features in audio signals, making them a promising alternative for SVC tasks. Chen et al. \cite{chen2024} introduced a method named LCM-SVC, which accelerates inference in latent diffusion models (LDM) for SVC through Latent Consistency Distillation (LCD), while retaining high sound quality and timbre similarity. Compared to traditional generative models, diffusion models better preserve the original sound details and musical characteristics. However, most existing diffusion-based SVC models employ a black-box structure, lacking intuitive interpretability and control over the generation process. This presents new challenges in optimizing and improving these models.

To address these challenges, this paper proposes a novel approach called The innovation of this study lies in the development of LHQ-SVC, a lightweight and CPU-optimized Singing Voice Conversion model. Unlike previous models that rely on GPU-heavy computations, LHQ-SVC is designed to run efficiently on CPUs while maintaining high-quality audio conversion. Key innovations include:
\begin{itemize}
    \item Model Size Reduction: LHQ-SVC significantly reduces the model size without compromising performance, making it suitable for deployment on devices with limited computational resources.
    \item CPU Optimization: By converting GPU-centric code to CPU-compatible operations and leveraging optimization libraries such as Intel MKL and OpenMP, LHQ-SVC achieves efficient parallel processing on multi-core CPUs.
    \item Improved Sampling Mechanism: We optimize the diffusion process by reducing the number of steps required in the conversion, enabling faster inference while maintaining high-quality results. 
\end{itemize}

\section{Related Work}

\subsection{Voice Conversion}

Voice Conversion (VC) is a technique that modifies the voice of a source speaker to sound like that of a target speaker while preserving the linguistic content. In recent years, VC has seen remarkable advancements with applications spanning across areas such as speaker identity conversion, speech enhancement, and cross-lingual voice translation \cite{sisman2020}. As a specialized subset of VC, Singing Voice Conversion (SVC) focuses on transforming one singer’s voice into another while retaining the musical elements such as pitch, rhythm, and timbre. This presents additional challenges beyond traditional VC tasks, as SVC must handle not only the vocal characteristics but also the melodic and rhythmic aspects of singing. Recent developments in deep learning have brought new solutions to the SVC field. Early methods predominantly relied on statistical models like Hidden Markov Models (HMMs) and Gaussian Mixture Models (GMMs). However, these models often struggled with issues related to the loss of audio quality and the need for parallel data, limiting their generalization capabilities \cite{kobayashi2014}. The shift towards neural network-based approaches has significantly improved both the quality and flexibility of voice and singing conversions.

While significant progress has been made in SVC through the introduction of deep learning methods such as GANs, diffusion models, and zero-shot techniques, challenges remain. Key issues include improving inference stability, reducing computational demands, and achieving better control over individual vocal timbres.

\subsection{Diffusion Model Applications}

Diffusion models, a class of generative models, operate by first adding noise to data in the forward process, and then reconstructing the data structure during the reverse process to generate samples \cite{yang2023}. These models have recently been introduced in the field of Singing Voice Conversion (SVC) to address the limitations of traditional methods when dealing with non-parallel data \cite{xue2024singvisio}. The core principle of diffusion models is to decompose the complex task of generating high-dimensional data into multiple simpler steps by progressively adding noise and learning the denoising process. Further advancing controllability in SVC, Wang et al. \cite{wang2024prompt} introduced Prompt-Singer, a method that allows users to control singer attributes such as gender, vocal range, and volume through natural language prompts. Although this method improves style control, the model's large size limits its scalability. Chen et al. \cite{chen2024ldmsvc} developed LDM-SVC, a zero-shot any-to-any SVC method based on latent diffusion models. LDM-SVC achieves faster training speed, but its inference quality is unstable.

Over and above these work, Lu et al. \cite{lu2024comosvc} proposed CoMoSVC, a consistency model-based SVC method designed to balance high-quality generation with fast sampling. The use of diffusion models in SVC has demonstrated great potential for generating high-quality audio and enabling fast inference, particularly in scenarios involving non-parallel data and complex timbre control. However, key challenges remain, including improving model efficiency, stability, and adaptability for real-world applications.

\section{Methodology}

\subsection{Model structure optimization and dynamic adjustment}
We propose a new model, LHQ-SVC, designed to significantly improve performance and reduce the model size while maintaining high accuracy, so that it can efficiently run on CPUs. The LHQ-SVC model focuses on optimising CPU execution, in contrast to the traditionally GPU-optimised code which relies heavily on parallel computation.

While CPUs have fewer cores compared to GPUs, their multi-core nature still allows for parallelism, albeit to a lesser extent. Using parallel frameworks like OpenMP or Intel Threading Building Blocks (TBB) can enable efficient multi-threaded programming, maximizing the CPU's multi-core capabilities. CPU performance tuning tools like Intel VTune Profiler can be used to identify performance bottlenecks, particularly in hotspot areas of code, such as I/O operations, memory access patterns, and cache management. Optimising these areas contributes to improving computational efficiency. In the training phase of LHQ-SVC, once robust training reaches a certain number of iterations, the network undergoes evaluation. Specifically, the performance metric, calculated as the ratio of the CLIP score to latency change before and after removing specific modules, is used to determine which parts of the model should be retained or removed. If the network's latency exceeds the target, modules that contribute to latency but not performance are removed. In contrast, modules with the highest performance contribution are duplicated to further strengthen the network's structure.

\begin{equation}
{\operatorname*{\hat{\epsilon}_\theta}}(t,\mathbf{z}_t)=\prod\left\{p(\text{Cross-Attention}(\mathbf{z}_t,\mathbf{c}),I),p(D(\mathbf{z}_t,t),I)\right\}
\end{equation}

This dynamic tuning process allows continuous robust training, leading to an optimised network structure without the need for retraining. The result is a well-balanced network with high performance and computational efficiency, suitable for deployment on CPU-based systems.

\subsection{Sampling mechanism optimization and efficient reasoning.}

\begin{figure}[htbp]
\centering
\includegraphics[width=.5\textwidth]{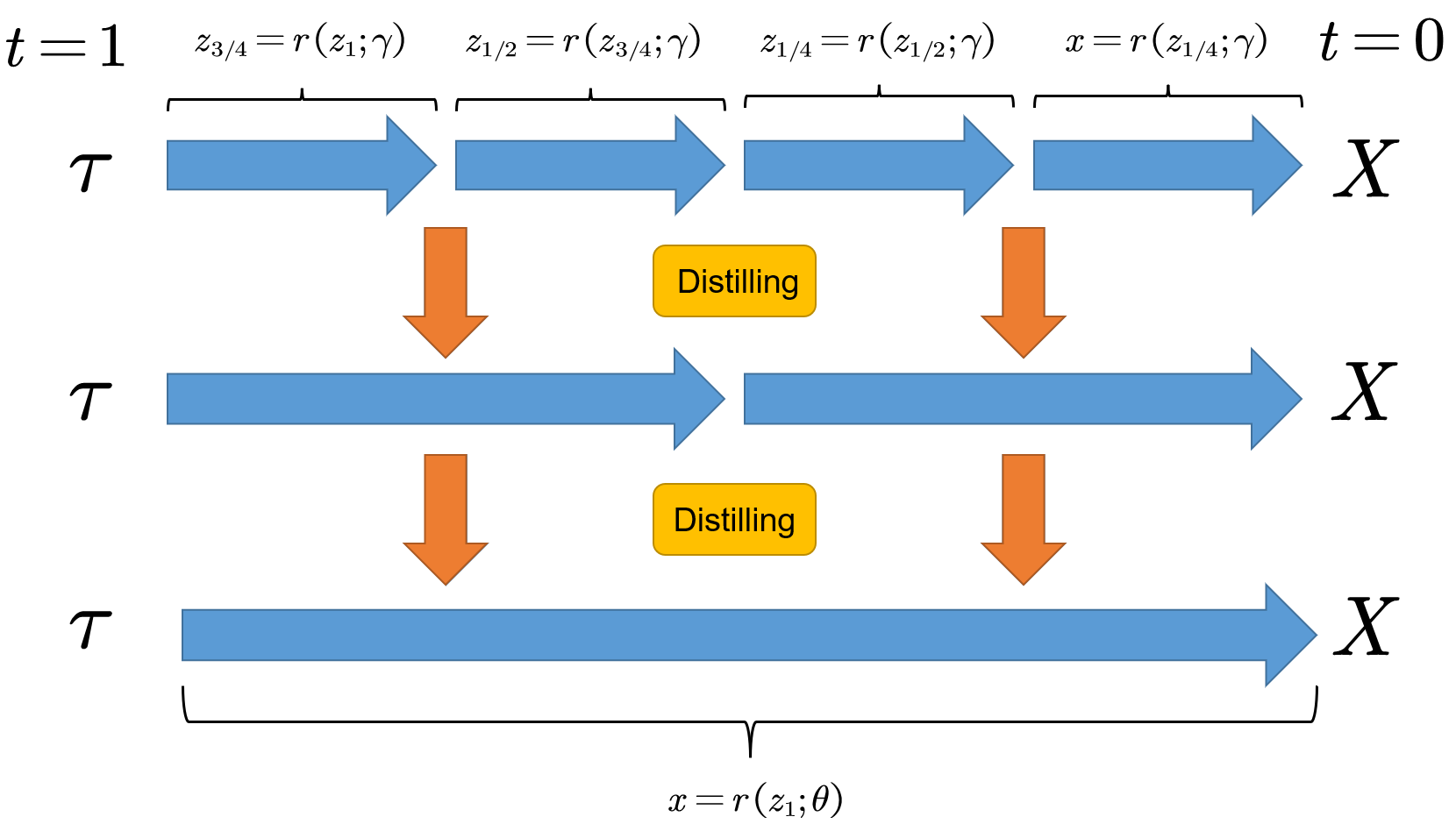} 
\caption{Progressive distillation diagram for diffusion modelling}
\label{frame1}
\end{figure}

Moreover, the time consumption of Mini-SVC can be further reduced by minimizing diffusion steps during the transition from the teacher model to the student model. By distilling the teacher model’s multi-step output into a single-step output for the student model, as shown in Figure ~\ref{frame1}, we effectively streamline the transformation process.

We consider a continuous-time diffusion framework where the training data is represented as $\tilde{x} \sim p(\tilde{x})$. The diffusion model operates with latent states denoted by $\tilde{y} = \{\tilde{y}_\tau \mid \tau \in [0,1]\}$, governed by a noise schedule determined by a differentiable function $\beta_\tau$. This function controls the signal-to-noise ratio (SNR), defined as $\theta_{\tau} = \log\left(\frac{\beta_\tau^2}{\gamma_\tau^2}\right)$, which monotonically decreases over time. The forward diffusion process $r(\tilde{y} \mid \tilde{x})$ is modeled as a Gaussian distribution, with the following Markovian structure:
\vspace{-8pt}
\begin{align}
r(\tilde{y}_\tau \mid \tilde{x}) &= \mathcal{N}(\tilde{y}_\tau; \beta_\tau \tilde{x}, \gamma_\tau^2 \mathbf{I}) \\
r(\tilde{y}_\tau \mid \tilde{y}_\rho) &= \mathcal{N}(\tilde{y}_\tau; \frac{\beta_\tau}{\beta_\rho} \tilde{y}_\rho, \gamma_{\tau|\rho}^2 \mathbf{I}) \\
\gamma_{\tau|\rho}^2 &= \left(1 - e^{\theta_\tau - \theta_\rho}\right) \gamma_\tau^2
\end{align}

The goal of the diffusion model is to recast the denoising task into an estimation problem, where $\tilde{x}_\psi(\tilde{y}_\tau) \approx \tilde{x}$. The model is trained using a weighted mean squared error loss function over uniformly sampled time points $\tau \in [0,1]$. This loss can be interpreted as either a variational lower bound on the log-likelihood of the data or as a type of denoising score matching:
\begin{equation}
\mathbb{E}_{\zeta, \tau} \left[v(\theta_\tau) \|\tilde{x}_\psi(\tilde{y}_\tau) - \tilde{x}\|_2^2 \right]
\end{equation}

After training, the model generates samples through various methods. The simplest of these is discrete-time ancestral sampling, derived by reversing the forward diffusion process:
\vspace{-8pt}
\begin{equation}
r(\tilde{y}_\rho \mid \tilde{y}_\tau, \tilde{x}) = \mathcal{N}(\tilde{y}_\rho; \nu_{\rho|\tau}(\tilde{y}_\tau, \tilde{x}), \gamma_{\rho|\tau}^2 \mathbf{I})
\end{equation}

where
\vspace{-8pt}
\begin{align}
\nu_{\rho|\tau}(\tilde{y}_\tau, \tilde{x}) &= e^{\theta_\tau - \theta_\rho} \frac{\beta_\rho}{\beta_\tau} \tilde{y}_\tau + \left(1 - e^{\theta_\tau - \theta_\rho}\right) \beta_\rho \tilde{x} \\
\gamma_{\rho|\tau}^2 &= \left(1 - e^{\theta_\tau - \theta_\rho}\right) \gamma_\rho^2
\end{align}

Using this reverse process, we define an ancestral sampling method. Starting from $\tilde{y}_1 \sim \mathcal{N}(\mathbf{0}, \mathbf{I})$, the ancestral sampling update follows:
\vspace{-8pt}
\begin{equation}
\begin{split}
\tilde{y}_\rho &= \nu_{\rho|\tau}\left(\tilde{y}_\tau, \hat{\tilde{x}}_\psi(\tilde{y}_\tau)\right) + \sqrt{\left(\gamma_{\rho|\tau}^2\right)^{1-\kappa} \left(\gamma_{\tau|\rho}^2\right)^{\kappa}} \zeta \\
&= e^{\theta_\tau - \theta_\rho} \frac{\beta_\rho}{\beta_\tau} \tilde{y}_\tau + \left(1 - e^{\theta_\tau - \theta_\rho}\right) \beta_\rho \hat{\tilde{x}}_\psi(\tilde{y}_\tau) \\
&+ \sqrt{\left(\gamma_{\rho|\tau}^2\right)^{1-\kappa} \left(\gamma_{\tau|\rho}^2\right)^{\kappa}} \zeta
\end{split}
\end{equation}

where $\zeta$ represents Gaussian noise, and $\kappa$ is a hyperparameter controlling the interpolation between noise levels. The DDIM (Denoising Diffusion Implicit Models) sampler, an efficient variant, solves an ODE to approximate the sampling process:
\vspace{-8pt}
\begin{equation}
\tilde{y}_\rho = \beta_\rho \tilde{x}_\psi(\tilde{y}_\tau) + \gamma_\rho \frac{\tilde{y}_\tau - \beta_\tau \tilde{x}_\psi(\tilde{y}_\tau)}{\gamma_\tau}
\end{equation}

This sampler provides superior performance to the standard ODE-based methods, particularly when $\tilde{x}_\psi(\tilde{y}_\tau)$ satisfies smoothness conditions. As the number of integration steps $N \rightarrow \infty$, the numerical error vanishes, enabling a trade-off between sample quality and computational efficiency.

\section{Experiments}

\subsection{Dataset Preparation}

\textbf{Singing and Speech Datasets} The datasets used in this study include the SVC-2023 Dataset \cite{huang2023singingvoiceconversionchallenge} and the DAMP (Data for Analysis of Musical Performances) Dataset \cite{8461660}. The SVC-2023 Dataset, widely used for benchmarking singing voice conversion models, provides parallel and non-parallel singing data across multiple languages, while the DAMP dataset offers a vast collection of over 30,000 singing performances collected from amateur singers using the Smule social singing app.

\textbf{Vocal Separation} To extract clean vocal tracks from the songs, we employed the \textit{Ultimate Vocal Remover} project. This involved two main steps: first, we used MDX-Net \cite{mdxnet2021} to remove the instrumental accompaniment from the audio files. Next, we applied the 6\_HP-Karaoke-UVR model \cite{uvr2021} to remove background harmonics from the vocal-separated files. The output was clean, isolated vocals that could be used for further processing and analysis.

\textbf{Audio Slicing} After obtaining clean vocal tracks, we sliced the audio into segments of less than 30 seconds in duration, manually discarding segments without vocals to ensure only those containing vocal content were retained. This step was automated using the open-source \textit{Audio Slicer} tool \cite{audioslicer2021}, which detects silent audio segments and slices the audio based on pitch, ensuring that the resulting audio clips primarily contained vocal content.

\textbf{Data Augmentation} All audio clips were resampled to a uniform rate of 40kHz and normalized. Data augmentation was applied to the audio clips by segmenting the dataset into clips of varying sizes, with partial overlaps between adjacent clips to increase diversity in the training set. This augmentation helps improve model robustness to different voice and singing patterns.

\subsection{Training Process}

\textbf{Feature Extraction} Before training, all audio files were resampled to 40kHz and normalized. We then extracted features using DIO for fundamental frequency (F0) and the ContentVec model to extract content-related features from the 12th layer, including F0 curves and voicing flags. These features were projected into a 256-dimensional space and concatenated to form the conditioning input for the decoding phase. We used a pre-trained vocoder \cite{vocoder2020} fine-tuned for singing voice synthesis and calculated Mel-spectrograms with a 512-point Fast Fourier Transform (FFT), a 512-point window size, and a 128-point hop size, producing 80 frequency bins. To evaluate the performance of the proposed singing voice conversion model, we compare it against several state-of-the-art models based on important metrics such as model size (MB), Short-Time Objective Intelligibility (STOI), Perceptual Evaluation of Speech Quality (PESQ), and Mean Opinion Scores (MOS) for both naturalness and similarity. The following methods are considered: RVC \cite{RVC2025}, SoVITS 4.0 \cite{svc_develop_team_so_vits_svc_2025}, SoVITS 5.0 \cite{ouor_sovits_svc_5_0}, Learn2Sing 2.0, \cite{xue2022learn2sing}, CoMoSVC \cite{lu2024comosvc}.

\textbf{Model Training} All models were trained using a single NVIDIA GTX 4090 GPU. The batch size was set to 48, and the learning rates were 1e-4 and 5e-5, using the Adam optimizer. We first conducted reconstruction experiments to evaluate the capability of different decoding phases in the autoencoder setup. Following this, we conducted two sets of experiments on the target singer’s dataset to fine-tune the model.

Additionally, we incorporated sampling steps for LHQ-SVC to assess the impact of different sampling methods on the quality of voice conversion. Conversion experiments were performed to evaluate the effect of these sampling steps on the overall conversion quality.

\subsection{Experimental Models and Results}

The performance of LHQ-SVC, along with several baseline models, is summarized in Table~\ref{tab1}. The results for LHQ-SVC and LHQ-SVC$_{mobile}$ stand out due to their lightweight nature and performance efficiency, especially for mobile device applications. LHQ-SVC, with a model size of 67.89 MB, achieves impressive results across multiple metrics, including the highest PESQ score of 3.02, and a Naturalness MOS of 4.02, which is the best among all models compared. These results indicate that LHQ-SVC excels in generating high-quality singing voice conversions while maintaining a relatively small model size. On the other hand, LHQ-SVC$_{mobile}$ is an even more compact version, with a model size of only 53 MB, optimized for mobile platforms. Although it shows slightly lower performance in terms of STOI and PESQ compared to LHQ-SVC, it still provides competitive results, especially in terms of Similarity MOS with a score of 3.91, making it a viable choice for lightweight applications requiring fast and efficient voice conversion on mobile devices. The balance between the performance and model size makes both LHQ-SVC and LHQ-SVC$_{mobile}$ particularly appealing for practical use, as they deliver strong results in Naturalness MOS and Similarity MOS while being resource-efficient.

\vspace{-10pt}
\begin{table}[h]
\centering
\caption{Comparison of Model Performance Metrics}
\label{tab1}
\setlength{\tabcolsep}{2.5pt} 
\begin{tabular}{@{}lccccc@{}}
\toprule
Model & Size (MB) & STOI ($\uparrow$) & PESQ ($\uparrow$) & N.MOS ($\uparrow$) & S.MOS ($\uparrow$) \\
\midrule
RVC & 102.66 & 0.70 & 2.97 & 3.99 & 3.96 \\
SoVITS 5.0 & 220.98 & \textbf{0.76} & 2.82 & 3.95 & \textbf{4.15} \\
SoVITS 4.0 & 620.13 & 0.40 & 2.12 & 3.12 & 3.47 \\
Learn2Sing 2.0 & 200.43 & 0.69 & 2.76 & 3.85 & 3.89 \\
CoMoSVC & 380.65 & 0.42 & 2.84 & 3.90 & 3.90 \\
LHQ-SVC & 67.89 & 0.72 & \textbf{3.02} & \textbf{4.02} & 4.10 \\
LHQ-SVC$_{mobile}$ & \textbf{53.24} & 0.63 & 2.88 & 3.85 & 3.91 \\
\bottomrule
\end{tabular}
\vspace{-10pt}
\end{table}

\begin{figure}[htbp!]
    \centering
    \includegraphics[width=0.5\textwidth]{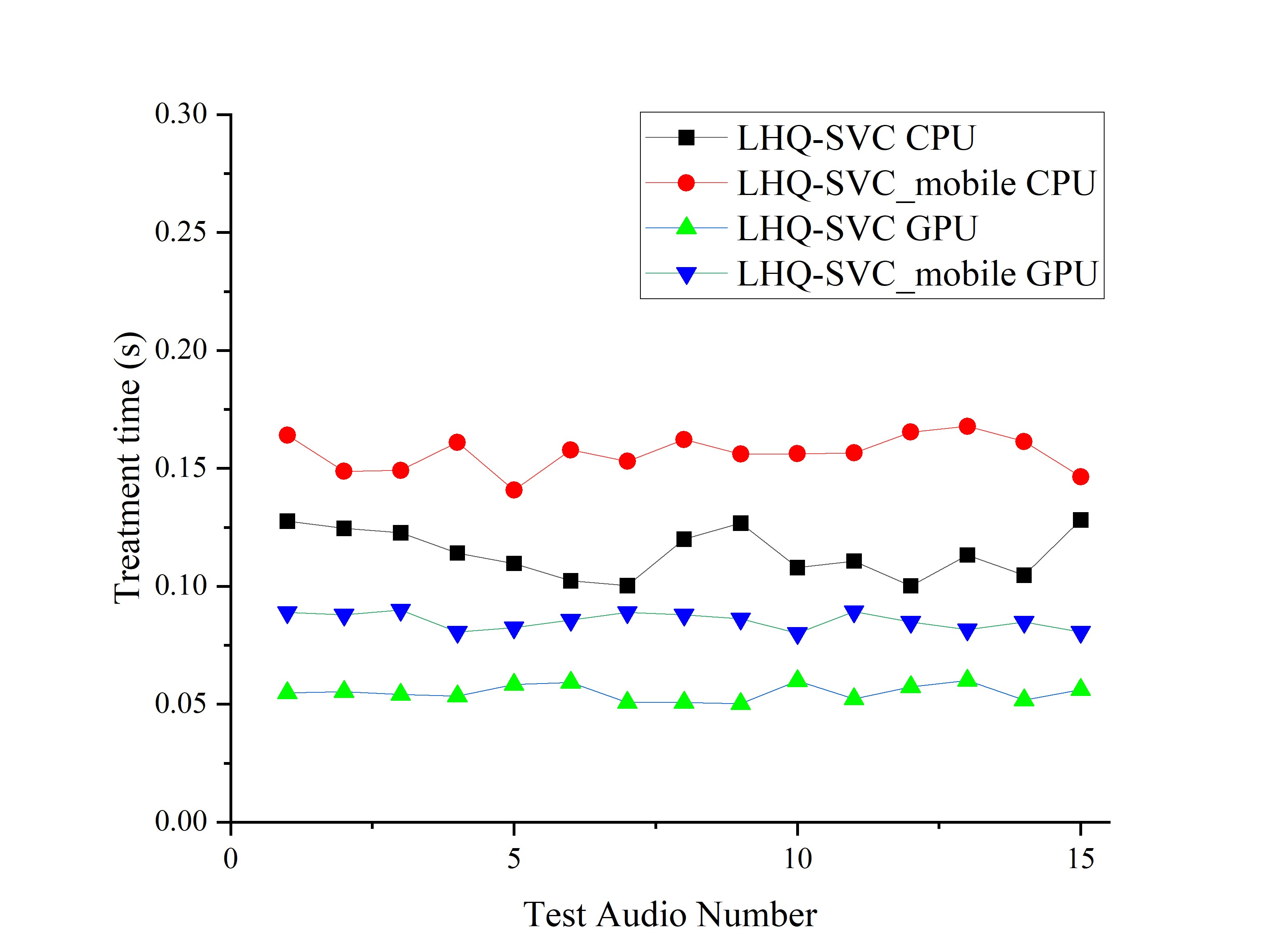}
    \vspace{-20pt}
    \caption{Comparison of GPU and CPU inference time on computer and cell phones}
    \label{re1}
\end{figure}

In order to verify that our LHQ-SVC is efficient enough, we ran it on the Intel Core i7-11370H on the computer side as well as on the Qualcomm Snapdragon 8 Gen3 processor equipped with the Qualcomm Snapdragon 8 Gen3 processor, and the results of the run are shown in the figure~\ref{re1}. As can be seen from the figure~\ref{re1}, after optimization, the processing speed has been greatly improved, in the Intel Core i7 model device reasoning 3min30s audio, the time from 1min (processing timespeed of about 60/210, about 30\% of the original audio time) to about 11s (processing timespeed of about 11/210, about 5.2\% of the original audio time), in the GPU device reasoning even more to about 9s (processing timespeed of about 6/210, about 2.2\% of the original audio time), to achieve the real-time efficient demand.), and the inference on the GPU device even reaches about 9s (the processing speed is about 6/210, which is about 2.2\% of the original audio time), which achieves the real-time and high efficiency requirements. At the same time, we simplified the algorithm running on a cell phone to run slightly slower than the client on a Qualcomm Snapdragon 8 Gen3, meeting the need for lightweight efficiency.

\vspace{-10pt}

\vspace{-10pt}

\section{Conclusions}
In this paper, we introduced LHQ-SVC, a novel lightweight Singing Voice Conversion model designed to run efficiently on CPUs while maintaining high-quality voice conversion performance. Through careful optimization, including the conversion of GPU-specific code and the use of parallel computing frameworks, we reduced the computational demand and model size significantly. Experimental results demonstrated that LHQ-SVC achieves competitive performance in terms of intelligibility, perceptual speech quality, and naturalness, while also delivering substantial improvements in processing speed across different hardware platforms.


\end{document}